\documentclass[reprint,aps,prd,nofootinbib,twocolumn,superscriptaddress,preprintnumbers]{revtex4-2}

\usepackage[T1]{fontenc}
\usepackage[usenames,dvipsnames]{xcolor}
\usepackage{aas_macros,hyperref,mathrsfs,orcidlink}
\hypersetup{colorlinks=true,citecolor=red,linkcolor=red,urlcolor=blue}

\usepackage{ulem}

\usepackage{graphicx,physics,amssymb,amsmath,amsthm,amsfonts,times}

\usepackage{amsmath}
\usepackage{enumitem}
\usepackage{esint}

\newcommand{\be}{\begin{equation}}
\newcommand{\ee}{\end{equation}}
\newcommand{\bea}{\setlength\arraycolsep{2pt} \begin{eqnarray}}
\newcommand{\eea}{\end{eqnarray}}

\def\b{\beta}

\def\d{\delta}
\def\D{\Delta}
\def\f{\frac}

\def\lm{\lambda}

\def\m{\mu} 
\def\n{\nu} 

\def\pl{\partial}

\def\t{\theta}

\begin{document}

\title{Probing the Scalar Hair of Rotating Horndeski Black Holes through Thick Disk Images}

\author{Qian Wan}
\affiliation{School of Physics and Optoelectronics, South China University of Technology, Guangzhou 510641, P. R. China}

\author{Yehui Hou}
\email{Corresponding author: yehuihou@sjtu.edu.cn}
\affiliation{Tsung-Dao Lee Institute, Shanghai Jiao-Tong University, Shanghai 201210, P. R. China}

\author{Minyong Guo}
\affiliation{School of Physics and Astronomy, Beijing Normal University, Beijing 100875, P. R. China}
\affiliation{Key Laboratory of Multiscale Spin Physics (Ministry of Education), Beijing Normal University, Beijing 100875,  P. R. China}

\begin{abstract}
Horizon-scale images of black holes provide a potential probe of fundamental physics, including tests of gravity and black hole hair. To assess the impact of scalar hair on accretion-flow imaging self-consistently, we construct an analytical model of a geometrically thick, magnetized disk around a rotating hairy black hole in Horndeski theory and analyze its 230 GHz image morphology. We find that scalar hair modestly alters the inflow and magnetic-field structure but strengthens gravitational redshift, markedly reducing the total flux and lensed ring brightness through relativistic transfer and spectral-shift effects. Moreover, we identify a novel, spatially resolved imaging feature: the maximum interferometric diameter of the first photon ring responds strongly to the hair parameter but shows little dependence on accretion-flow details, making it a promising observable for constraining black hole hair with future space-based interferometry.
\end{abstract}

\maketitle

\section{Introduction}
\label{sec1}

Black holes are natural laboratories for testing gravitational theories and potential departures from the no-hair paradigm. In general relativity (GR), stationary black holes are completely specified by their mass, spin, and charge \cite{Israel:1967za, Carter1971, Robinson:1975bv},  motivating the exploration of extensions that introduce additional degrees of freedom while maintaining theoretical consistency.
Nonminimal couplings or special dynamical fields can endow black holes with nontrivial scalar configurations \cite{Bekenstein:1974sf, Sotiriou:2011dz, Herdeiro:2014goa}, which often concentrate near the horizon and can impact accretion dynamics and photon trajectories \cite{Cunha:2015yba}.  
These effects may leave observable imprints in black hole images, which could be identified by next-generation very-long-baseline interferometry (VLBI) facilities such as the next-generation EHT  \cite{Ayzenberg:2023hfw, Johnson:2023ynn, Tiede:2022grp} and the Black Hole Explorer (BHEX) \cite{Johnson:2024ttr, Lupsasca:2024xhq}. In particular, BHEX will provide the angular resolution needed to probe the strongly lensed rings, which are especially sensitive to the underlying spacetime geometry.

As the most general four-dimensional scalar-tensor theory with diffeomorphism invariance and second-order field equations,  Horndeski theory provides a broad setting in which scalar hair may arise \cite{Horndeski:1974wa, Kobayashi:2011nu}. The theory is also widely employed in cosmology and astrophysics \cite{Gleyzes:2014dya, Bellini:2014fua, Kobayashi:2019hrl}, combining theoretical generality with observational relevance.
Recently, a spherically symmetric black hole solution with asymptotically flat behavior and logarithmic corrections sourced by scalar hair was obtained in \cite{Bergliaffa:2021diw}. Its rotating counterpart was later constructed via the Newman-Janis algorithm \cite{Walia:2021emv} and has since been explored in various contexts \cite{Seifi:2022xyw,Jha:2022tdl,Rayimbaev:2023bjs,Donmez:2024gqb,Ashraf:2024xwm,Wu:2025ccc,Zhen:2025nah,Yang:2023lcm,Lei:2023wlt,Donmez:2024lfi,Yang:2024cjf,Carvajal:2025emj}, including gravitational lensing \cite{Atamurotov:2022slw,Bessa:2023ykd}, black hole images \cite{Wang:2023vcv,Hu:2023pyd,Gao:2023mjb,Heydari-Fard:2023kgf,Sucu:2025vwy,Meng:2025ivb,Zeng:2025pch,Heydari-Fard:2025biz}, and polarization signatures under simplified emission models \cite{Shi:2024bpm,Chen:2025ysv}.
Nonetheless, studies incorporating horizon-scale accretion flows and the corresponding imaging features remain limited.

As the central engines of many low-luminosity active galactic nuclei, supermassive black holes are typically surrounded by magnetized, advection-dominated accretion flows (ADAFs) \cite{Narayan:1994xi, Yuan:2014gma, Blandford:1998qn, Balbus:1991ay}, where the plasma exhibits pronounced radial inflow and inefficient Coulomb coupling between ions and electrons \cite{Narayan:1995ic, Mahadevan:1996jf, Quataert:1998yn, Yuan:2003dc}. 
General-relativistic magnetohydrodynamic (GRMHD) simulations, though have highly computational demands, offer the most self-consistent approach currently to studying these accretion flows \cite{EventHorizonTelescope:2019pcy}, and can be extended to explore dynamics in modified-gravity spacetimes \cite{Jiang:2023img, Dihingia:2024cch, Uniyal:2025hik, Uniyal:2025uvc}. 
At the same time, semi-analytic accretion models remain an important complementary tool, enabling rapid parameter exploration and controlled studies aimed at isolating individual physical effects \cite{Michel:1972oeq, Page:1974he, Popham:1998iw, 2000ApJ...534..734M, Broderick:2005my}. 
However, many formulations in this class rely on simplifying assumptions---such as self‑similar radial structure, geometric thinness, or nearly circular orbital motion---that are not generally valid in the strongly relativistic, horizon-proximate regime.

Near the event horizon, gravity dominates the dynamics, making the ballistic approximation---treating the plasma motion as timelike geodesics---an effective and analytically tractable description of the flow \cite{Ulrich:1976zz, Tejeda:2012kb, Tejeda:2019lie}. 
Using this approach, \cite{Hou:2023bep} constructed a self-consistent model in Kerr spacetime, obtaining closed-form expressions for the density, temperature, and magnetic field by exploiting the separability of the geodesic equations \cite{Carter:1968rr}. 
Building on this framework, we extend the analysis to rotating Horndeski black holes and investigate the associated implications for black hole imaging.
We begin by examining how the scalar-hair parameter modulates key flow properties, including the density profile, thermal structure, and magnetic-field configuration. 
We then analyze the resulting observational signatures, such as the spatially resolved intensity maps, total flux, and the morphologies of the critical curve and inner shadow associated with this accretion flow. Particular attention is devoted to the influence of the hair parameter on the maximal interferometric diameter of the first photon ring \cite{Farah2025}, with emphasis on regimes that remain potentially accessible to BHEX observations.

The remainder of this paper is organized as follows.
Sec.~\ref{sec2} reviews the rotating hairy black hole solution in Horndeski theory and outlines the geodesic equations.
Sec.~\ref{sec3} describes the accretion-flow model adopted in this work, including the density, temperature, and magnetic-field structures.
Sec.~\ref{sec4} presents a detailed analysis of the corresponding images and their observable features under various spacetime and flow parameters, while Sec.~\ref{sec5} specifically examines the constraining power of the first photon ring on black hole scalar hair.
Sec.~\ref{sec6} offers our concluding remarks.
Throughout this paper, we adopt geometrized units with $G = c = 1$.

\section{Rotating Horndeski Black Holes}
\label{sec2}

In this section, we briefly review the rotating hairy black hole solution in Horndeski theory and discuss the corresponding geodesic motion, which is essential for constructing the analytical disk model presented in the next section.

\subsection{The spacetime}

The rotating metric under study is obtained from the spherically symmetric solution in Horndeski theory via the Newman-Janis algorithm \cite{Walia:2021emv}.
In Boyer-Lindquist (BL) coordinates, it takes the form
\begin{eqnarray}
	\begin{aligned}
		\label{metric}
		\dd s^{2}=&-\left(\frac{\Delta-a^2\sin^2\theta}{\Sigma}\right)\dd t^2+\frac{\Sigma}{\Delta}\dd r^2+\Sigma\dd\theta^2 \\
		&+\frac{2a\sin^2\theta}{\Sigma}\left[\Delta-(r^2+a^2)\right]\dd t\dd\phi \\
		&+\frac{\sin^{2}\theta}{\Sigma}\left[(r^{2}+a^{2})^{2}-\Delta a^{2}\sin^{2}\theta\right]\dd\phi^{2},
	\end{aligned}
\end{eqnarray}
where $\Delta=r^2+a^2-2Mr+hr\ln(r/2M)$, $\Sigma=r^2+a^2\cos^2\theta$. 
The spacetime is characterized by three parameters: the black-hole mass parameter $M$, spin parameter $a$ and scalar-hair parameter $h$. 
The hair parameter quantifies the deviation from the Kerr geometry. 
In the limit $h \to 0$, Eq.~\eqref{metric} recovers the Kerr metric, whereas for $a \to 0$, it reduces to the static spherical solution in Horndeski theory \cite{Bergliaffa:2021diw}.

The spacetimes contain a scalar field with a density profile $\rho(r) \sim r^{-3}$, 
which contributes an additional gravitational potential around the black hole, leading to
\begin{equation}
	\label{Phieff}
	\Phi \sim  -\f{M}{r}  + \frac{h}{2r}\ln{\left(\f{r}{2M}\right)} \,,
\end{equation}
where the first term corresponds to the central black-hole contribution, while the second term arises from the scalar field.
Analysis of $\Phi(r)$ reveals the existence of a critical radius $r = 2 \,\text{e}M \approx 5.4 M$, where $\text{e}$ denotes the base of the natural logarithm, at which the force exerted by the scalar field on test particles vanishes.
For $r > 2 \,\text{e}M$, the scalar field exerts an attractive (repulsive) force toward the black hole when $ h \lessgtr 0$, corresponding respectively to dark-matter-like and dark-energy-like configurations. In what follows, we focus on the physically relevant, dark-matter-like case $h \leq 0$. 

The metric in Eq.~\eqref{metric} features a Kerr-like ring singularity at $\Sigma=0$. Outside it, the condition $\Delta=0$ typically yields two real roots defining the Cauchy and event horizons, $r_+ \geq r_-$ \cite{Afrin:2021wlj}.  
Since $\D(2M) = a^2 \geq 0$, the event horizon lies inside it, $r_+\leq 2M$. As $h$ decreases, the outer horizon radius $r_+$ also shrinks. For a given spin $a$, there exists a critical value $h_c$ for which $r_+=r_-$, corresponding to the extremal configuration.
When $h < h_c$, the horizons disappear and the spacetime develops a naked singularity. For completeness, the regions of parameter space corresponding to black hole solutions and naked singularities are displayed in the left panel of Fig.~\ref{parah} in the Appendix. In this work, we focus exclusively on configurations in which the singularity is cloaked by an event horizon, consistent with the cosmic censorship conjecture \cite{penrose1969gravitational}. Accordingly, our analysis is restricted to the regime $h > h_c$ throughout.

The rotation of spacetime induces frame dragging.
Zero-angular-momentum observers (ZAMOs) outside the horizon rotate with a nonzero angular velocity $\omega=-g_{t\phi}/g_{\phi\phi}$, which increases monotonically toward the horizon and approaches the horizon angular velocity at $r = r_+$,
\begin{equation}
	\Omega_H=\frac{a}{r_+^2+a^2}\,.
\end{equation}
The spacetime rotation also gives rise to an ergosphere, defined by the surface $g_{tt} = 0$ at $r = r_\text{eg}(\theta)$.
Inside the ergoregion, the Killing vector $\partial_t$ becomes spacelike and frame dragging forces all particles to co-rotate with the black hole.
Energy extraction from this region can proceed through the Penrose process \cite{penrose1969gravitational}, the Blandford-Znajek mechanism \cite{Blandford:1977ds}, or its magnetohydrodynamic extension \cite{takahashi1990magnetohydrodynamic}, converting rotational energy into escaping particle or electromagnetic flux.

Notably, the spacetime in Eq.~\eqref{metric} resembles that of a black hole surrounded by perfect-fluid dark matter (PFDM) \cite{Li:2012zx, Das:2020yxw,Xu:2017bpz}, with the two being related by mass parameter redefinition. Hence, our analysis can be directly extended to rotating black holes embedded in PFDM, a system that has been the focus of several recent imaging studies \cite{Haroon:2018ryd, Hou:2018avu, Atamurotov:2021hck, Saurabh:2020zqg, Heydari-Fard:2022xhr, Aslam:2025hgl, Ban:2025uip}. 

\subsection{Geodesics}

The motion of a test particle in the given background spacetime can be conveniently analyzed using the Hamilton-Jacobi formalism. The Hamilton-Jacobi equation reads
\begin{equation}\label{HJeq}
	\partial_\lambda S + \mathcal{H} = 0 \,, \quad \mathcal{H}=\frac{1}{2}\,g^{\mu\nu}\pl_{\mu}S\, \pl_{\nu}S \,,
\end{equation}
where $S(x^{\m},\lambda)$ is the Hamilton's principal function, $\lambda$ is an affine parameter along the particle's worldline, and $\mathcal{H}$ denotes the covariant Hamiltonian for a freely moving particle. The associated four-momentum is $p_\mu= \pl_{\mu} S$.
Because the spacetime represented by Eq.~\eqref{metric} is stationary and axisymmetric, there exist two conserved quantities for test-particle motion $E=-p_t$ and $L=p_\phi$, corresponding to the energy and the axial component of angular momentum. Substituting these expressions into the Hamiltonian definition yields
\begin{equation}
	\label{Hform}
	\begin{aligned}
		&\bigg\{\Delta(r)\left(\pl_rS\right)^2 - \frac{[(r^2+a^2)E-aL]^2}{\Delta(r)}\bigg\} - 2\mathcal{H}r^2 \\= 
		&-\bigg\{\left(aE\sin\theta-\frac{L}{\sin\theta}\right)^2 + \left(\pl_{\t}S\right)^2  \bigg\}+ 2\mathcal{H}a^2\cos^2{\t}.
	\end{aligned}
\end{equation}
By imposing the on-shell condition $\mathcal{H} = -\mu^2/2$, where $\mu$ denotes the particle’s rest mass, Eq.~\eqref{Hform} implies that the Hamilton-Jacobi equation admits separability under the ansatz: $S =-Et+L\phi +S_r(r) + S_{\t}(\t)+\m^2\lm/2$. Substituting this form into Eq.~\eqref{HJeq}, one finds that the Hamilton-Jacobi equation separates into independent $r$ and $\t$ equations, yielding the following components of the four-momentum:
\begin{eqnarray}
	\label{eom}
		p_{\mu}\,\dd x^{\mu} = -E \,\dd t + L \,\dd \phi + \sigma_r\f{\sqrt{\mathcal{R}}}{\D}\,\dd r + \sigma_\theta\sqrt{\Theta} \,\dd\t \,.
\end{eqnarray}
Here, the symbols $\sigma_r,\sigma_\theta = \pm 1$ indicate the direction of motion of the test particle along the radial and polar coordinates, respectively.
The corresponding effective potentials in the radial and polar directions are given by
\begin{equation}
	\begin{aligned}
		&\mathcal{R}(r)=\left[(r^2+a^2)E-aL\right]^2-\Delta\left[\mu^2r^2+Q+(aE-L)^2\right],\\
		&\Theta(\theta)=Q+a^2(E^2-\mu^2)\cos^2\theta-L^2\cot^2\theta.
	\end{aligned}
\end{equation}
Here $Q$ is the Carter constant arising from the separability of the Hamilton-Jacobi equation \cite{Carter:1968rr}.
As in the Kerr spacetime, $Q$ is associated with a hidden (non-geometric) symmetry and provides the fundamental reason for the complete separability of Eq.~\eqref{HJeq}.

From the form of the effective potentials, the geodesic motion in the polar direction is identical to that in the Kerr spacetime.
In particular, the polar trajectories can be grouped into two distinct categories: (i) oscillatory orbits that repeatedly cross the equatorial plane and (ii) vortical orbits confined within conical shells outside the equatorial plane \cite{Hou:2023hto, Zhang:2023cuw}. 

In contrast, the radial motion becomes considerably more intricate when the hair parameter is nonzero, owing to the presence of a logarithmic term in the radial potential.
Nevertheless, by examining the local extrema of $R(r)$, one can still identify the key characteristics of particle dynamics, such as the existence of stable and unstable circular orbits and their associated conserved quantities \cite{Afrin:2021wlj}. 

In this work, we adopt a numerical ray-tracing scheme to integrate the null geodesics, thereby circumventing the need to manipulate the cumbersome analytic form of the potential. 
Hereafter we set $M = 1$ without loss of generality.

\section{The Disk Model}
\label{sec3}

At the horizon scale, the plasma inflow moves rapidly in both the toroidal and poloidal directions under the influence of strong gravity and frame dragging \cite{Ricarte:2022wpd}. 
To obtain explicit closed-form expressions for the number density, temperature, and magnetic field geometry of the accretion flow, we assume that the contributions of gas pressure $p_\text{gas}$ and magnetic pressure $p_B$ to the flow acceleration are negligible compared with gravity. 
Under this condition, the accretion flow moves along timelike geodesics, i.e., it satisfies the ballistic approximation  \cite{Ulrich:1976zz, Abramowicz:2011xu, Tejeda:2012kb}. 
\cite{Hou:2023bep} showed that under this condition the disk structure near the horizon becomes substantially simplified. In a matter-dominated disk with non-relativistic ions, both the sound speed and the Alfvén speed in the fluid-comoving frame remain sub-relativistic, making the ballistic approximation well justified \cite{Abramowicz:2011xu}.

Note that in magnetically arrested disks (MADs) of low-luminosity AGNs such as M87, both the plasma beta, $\beta=p_\text{gas}/p_B$, and the magnetization exhibit vertical stratification \cite{Tchekhovskoy:2011zx, McKinney:2012vh}. In the main emission region near the equatorial plane,  $\beta$ is typically of order unity or larger \cite{EventHorizonTelescope:2019ggy, EventHorizonTelescope:2022xqj}. Provided that the gas pressure does not play a dynamically dominant role, magnetic pressure effects can be treated as subdominant, and the streamline remains primarily gravity-driven. 
Under this assumption, the inflow in the disk region can be approximated as ballistic, and the streamlines may be modeled using timelike geodesics \cite{Huerta:2006sc, Tejeda:2012kb}.
In contrast, the jet and wind regions differ qualitatively from the disk. In jets, magnetic fields are dynamically important and $\b$ is small. In particular, in the nearly evacuated jet spine near the axis, where $\b \ll 1$, a force-free description is more appropriate in GRMHD \cite{Gralla:2014yja, Chael:2024gvx}. Nevertheless, jet emission-typically subdominant at millimeter wavelengths-is not included in the present work.

\subsection{Conical motion}

For computational convenience, we further constrain each streamline to lie on a conical surface with a fixed polar angle $\theta$. To maintain a stable vertical structure, the geodesic must be located at a local minimum of the angular effective potential \cite{Hou:2023bep}, i.e., $\Theta=\partial_\theta\Theta = 0$, and $\pl_{\t}^2\Theta \leq 0$. This condition imposes the following constraints on the angular momentum and the Carter constant: 
\begin{equation}
	\label{conEL}
	\begin{aligned}
		&L_\text{c} = \pm_L\,a\sin^2\t \sqrt{E^2-\mu^2}\,, \\
		&Q_\text{c} = -a^2\cos^4\t \,(E^2-\mu^2)\,,
	\end{aligned}
\end{equation}
where $\pm L$ denotes the sign of $L$. Clearly $L_c,Q_c$ are real only when $E\geq\mu$. Therefore, each stable, conical streamline is fully characterized by its polar angle $\t$ and the energy $E$, extending from infinity down to the event horizon $r = r_+$, which serves as the inner boundary of the accreting plasma. It should be noted, however, that the constraint of Eq.~\eqref{conEL} is required only for the thick disk where flows are off-equatorial  \cite{Hou:2023hto}; for an ideal thin disk  perfectly confined to the equatorial plane, it is not necessary.

We adopt the conical flow primarily for computational convenience in disk imaging \cite{Zhang:2024lsf}. Although this idealization may not strictly hold even for highly sub-Keplerian flows, it enables explicit, closed-form solutions for the disk structure, which are crucial for efficient parameter-space exploration and for investigating a wide range of physical scenarios.
From a phenomenological standpoint, two limiting flow configurations are commonly considered in black hole imaging studies: circular motion and free-fall motion. In this work, we adopt the free-fall limit with $E=\mu$. The differences between these two cases have been thoroughly examined in the literature (see, e.g., \cite{Vincent:2022fwj}), and we therefore expect qualitatively similar behavior in our setup. For this reason, we do not further investigate images produced by circular flows.

Moreover, in steady-state ideal GRMHD, the magnetic field is aligned with the fluid velocity in the poloidal plane, satisfying $B^\theta u^r=B^ru^\theta$. 
In the case of purely circular motion with $u^r=u^\theta=0$, this relation is trivially satisfied and therefore imposes no constraint on the magnetic-field. In contrast, when a nonzero radial velocity is present, the ideal-MHD condition provides a nontrivial constraint that helps determine the magnetic-field geometry.

\subsection{Density and temperature}

To derive the electron number density and temperature in the fluid-comoving frame (FCF), we assume that the distribution of horizon-scale electrons in the FCF is described by isotropic Maxwell-Jüttner function with dimensionless temperature $\Theta_\text{e}=k_\text{B}T_\text{e}/m_\text{e}$, where $k_\text{B}$ is the Boltzmann constant.
Solving the covariant continuity equation and adiabatic relations along streamlines yields explicit expressions for $n_\text{e}$ and $T_\text{e}$, which deviate from simple power-law scalings \cite{Hou:2023bep}

\begin{equation}
	\label{neeq}
	\begin{aligned}
		n_\text{e}(r,\theta)&=n_\text{e}(r_+,\theta)\sqrt{\frac{\mathcal{R}_\text{c}(r_+,\theta)}{\mathcal{R}_\text{c}(r,\theta)}}\,, \\ 
		\mathcal{R}_\text{c}(r,\theta) &\equiv \mathcal{R}(r)\big|_{L = L_\text{c}, Q = Q_\text{c}} \,, 
	\end{aligned}
\end{equation}
\begin{equation}
	T_\text{e}(r,\theta)=\left\{
	\begin{aligned}
		& T_\text{e}(r_+,\theta)\left[\frac{\mathcal{R}_\text{c}(r_+,\theta)}{\mathcal{R}_\text{c}(r,\theta)}\right]^{\frac{(1+z)}{3(2+z)}},\,\, \, \, z\ll z_\text{c} \\
		&T_\text{e}(r_+,\theta)\left[\frac{\mathcal{R}_\text{c}(r_+,\theta)}{\mathcal{R}_\text{c}(r,\theta)}\right]^{1/6},\,\,\,\,\, \quad z\gg z_\text{c}
	\end{aligned}\right. \label{Teeq}
\end{equation}
where $z=T_\text{ion}/T_\text{e}$ denotes the ion-to-electron temperature
ratio, assumed to remain constant near the horizon.
Values $z\ll  z_\text{c} \equiv m_\text{ion}/(k_\text{B}T_\text{e}(r_+,\theta))$ correspond to non-relativistic ions, whereas $z\gg z_\text{c}$ indicate relativistic ions.
For simplicity, we take the boundary electron temperature at the horizon, $T_\text{e}(r_+,\theta)=T_{+}$,  to be uniform and assume that the electron number density there follows a Gaussian profile, 
\begin{equation}
	n_\text{e}(r_+,\theta)=n_+\exp[-\left(\frac{\sin\theta-1}{\sigma}\right)^2],
\end{equation}
where $\sigma$ is the standard deviation.
In the following calculations, we adopt $T_+=10^{11}\,\mathrm{K}$ and $n_+ = 10^{6}\,\mathrm{cm}^{-3}$, representative of the disk emission region near M87* \cite{Vincent:2022fwj}.
The corresponding dimensionless electron temperature is $\Theta_\text{e}\approx 15$, and the critical ratio is $z_\text{c}\approx 122$.

\begin{figure}[htbp]
	\centering
	\includegraphics[width=2.9 in]{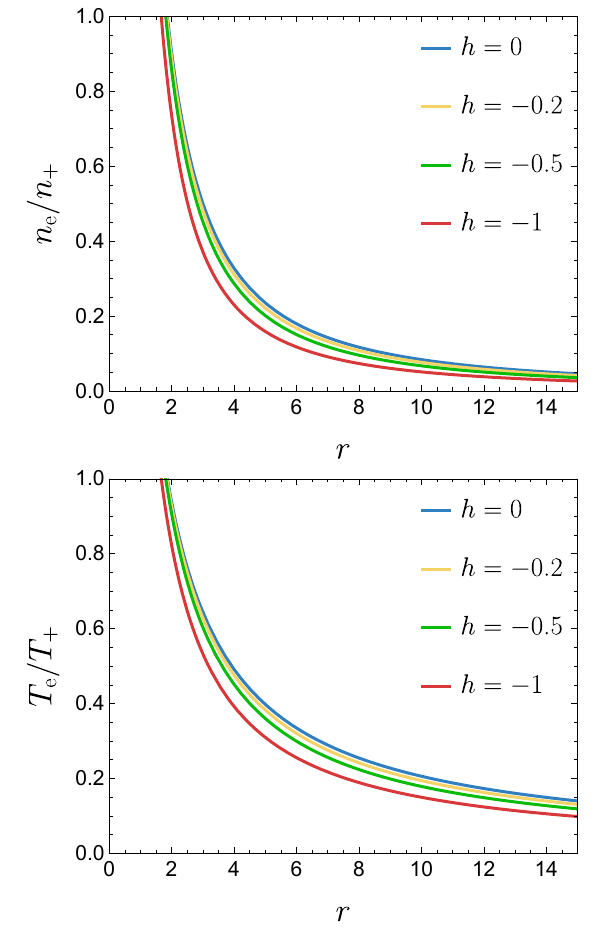}
	\caption{Radial profiles of the particle number density $n_\text{e}$ (top panel) and electron temperature $T_\text{e}$ (bottom panel) in the equatorial plane, where $n_+ = 10^6\,\mathrm{cm}^{-3}$ and $T_+ = 10^{11}\,\mathrm{K}$ represent the corresponding values at the event horizon $r_+$. The parameters are set to $a=0.5$, $z=20$, $\sigma=0.2$ and $E=\mu$. The unit of length is the gravitational radius $GM/c^2$.}
	\label{fignT}
\end{figure}

To preliminarily assess the effect of the hair parameter $h$ on the flow structure, Fig.\ref{fignT} presents the radial profiles of the electron number density and temperature in the equatorial plane for various $h$, with the spin fixed at $a=0.5$. 
Both quantities decrease rapidly toward zero with increasing distance from the black hole. The number density exhibits a steeper decline, following approximately $r^{-3/2}$ for free-fall flows and $r^{-2}$ in the other cases, whereas the temperature scales as $T_\text{e}\sim n^{\frac{2(1+z)}{3(2+z)}}\sim n^{2/3}$ for large $z$, \footnote{Here we assume $z \ll z_\text{c}$ and adopt the first relation in Eq.~\eqref{Teeq}. For general $z$, the expression for $n_\text{e}$ (Eq.~\eqref{neeq}) and the adiabatic relation between $T_\text{e}$ and $n_\text{e}$ remain valid, although the explicit form of $T_\text{e}$ becomes more complicated.} evaluated at large radii where $\mathcal{R}(r) \sim r^3$ for free-fall flows and $\mathcal{R}(r) \sim r^4$ otherwise. 
As the event horizon contracts with increasing $|h|$, both the $n_\text{e}$ and $T_\text{e}$ distributions shift slightly toward smaller radii.
Comparing the profiles for different $h$ shows that decreasing $h$ leads to a marginally faster decay of both $n_\text{e}$ and $T_\text{e}$; however, this effect is considerably weaker than that produced by variations in the flow parameters themselves \cite{Hou:2023bep}.

In addition, the hair parameter $h$ also affects the mass accretion rate of the disk, i.e., the flux of the covariant mass flow $\rho u^\mu$ through a closed surface. 
By selecting a closed sphere $\Sigma$ at a radius outside the horizon, we have $\dot{M}=-\oint_\Sigma\rho u^\mu\dd\Sigma_\mu$. 
For the conical flow, we further have
\begin{equation}
	\begin{aligned}
		\dot{M}=2\pi m_\text{ion}(r_+^2+a^2)\int_0^\pi n_\text{e}(r_+,\theta) \sin\theta\dd\theta\,.
	\end{aligned}
\end{equation} 
indicating that the parameters $a$, $h$ affect $\dot{M}$ solely through their influence on the horizon radius $r_+$. 
It is showed in Fig.~\ref{parah} that decreasing $h$ leads to a reduction in the event horizon radius $r_+$, which in turn lowers the accretion rate. Similarly, increasing $a$ decreases the factor $(r_+^2 + a^2)$, also resulting in a reduced accretion rate.
We note that the above feature depends on the specific assumptions of the disk model. In addition, since we consider a radiation-inefficient accretion flow, the radiated energy flux is treated as decoupled from the dynamics of the flow and therefore does not back-react on the accretion rate.

It should be noted that deriving the particle number-density distribution under ballistic approximation does not necessarily rely on the separability of the timelike geodesic equation.
For a steady, axisymmetric conical inflow, the continuity equation reduces to a one-dimensional radial form with an analytically solvable radial velocity.
In this case, an explicit expression for the number-density profile can still be obtained.
However, assessing the stability of such a conical configuration requires additional considerations.

\subsection{Frozen-in magnetic field}

\begin{figure*}[htbp]
	\centering
	\includegraphics[width=6.2in]{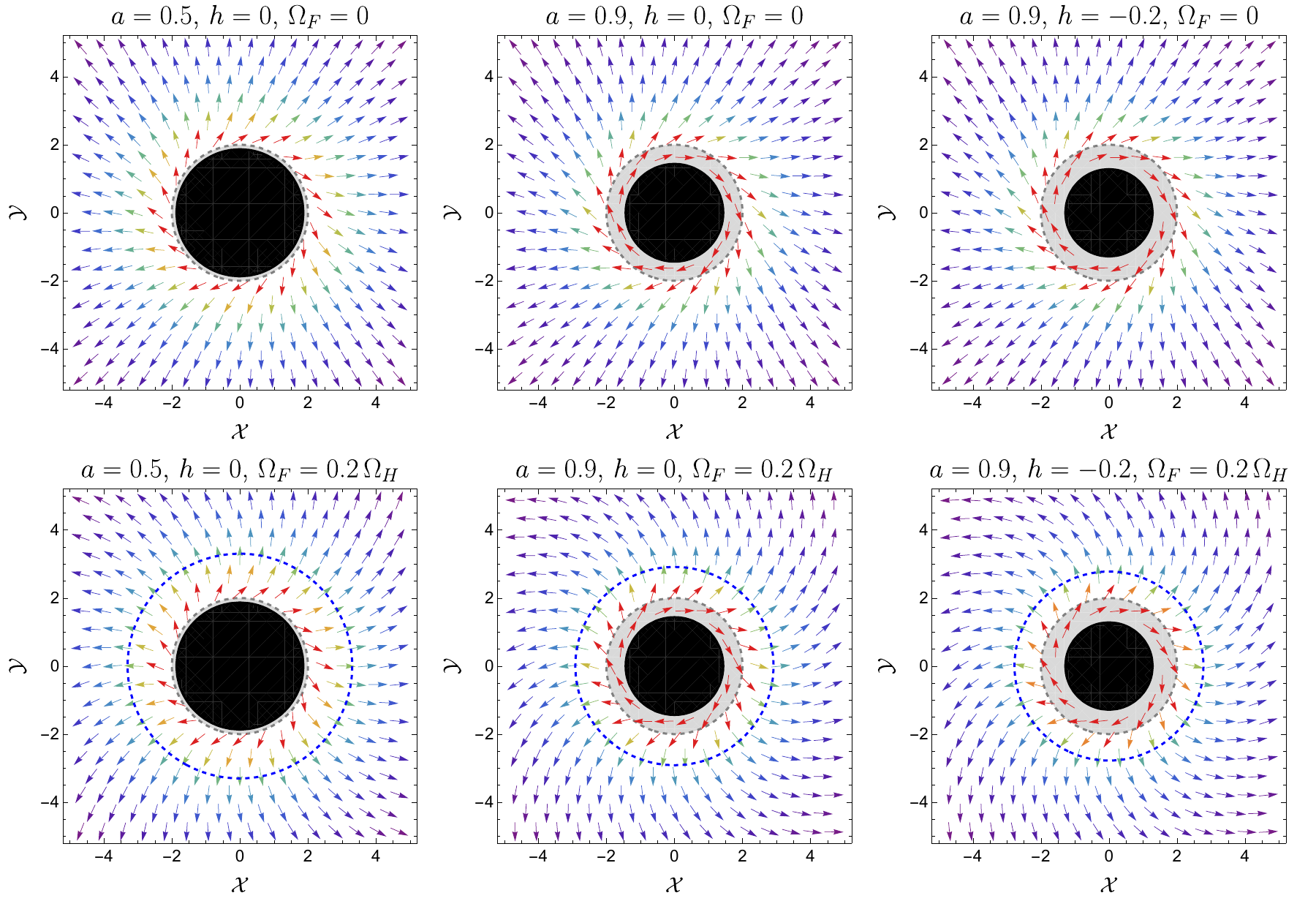}
	\caption{This figure depicts the equatorial plane structure of the magnetic field $B_\mu$ within a free-falling accretion flow with $E=\mu$ under two different parameter sets: $\Omega_F=0$ (top row) and $\Omega_F=0.2\,\Omega_H$ (bottom row). Here the black and gray regions represent the event horizon and the ergosphere, respectively, while the blue dashed line marks the critical radius $r_\text{c}$ at which magnetic field lines switch from retrograde to prograde rotation.}
	\label{Bmuplot}
\end{figure*}

To describe the electromagnetic field, we adopt the definitions $E^{\mu} = F^{t\mu}$, $B^{\mu} = -(*F)^{t\mu}$, the primitive variables commonly employed in GRMHD simulations. These quantities are not covariant and therefore do not transform as four-vectors. Specifically, $E^{\mu}, B^{\mu}$ differ from the physical electric and magnetic fields measured by a spacetime normal observer only by a lapse factor \cite{Komissarov:2004ms}. For terminological clarity and precision, we thus refer to them as ``pseudo'' fields.

In a steady-state, axisymmetric ideal-MHD configuration, the pseudo-magnetic field $B^\mu$ is aligned with the flow streamlines in the poloidal plane, satisfying $B^r u^{\theta}= B^{\theta}u^r$, where $u^{\mu}$ is the flow four-velocity. 
At this point, the electromagnetic field admits two conserved quantities along each streamline \cite{thorne1982electrodynamics}: the stream function $\psi = A_\phi$, which differs by a factor of $2\pi$ from the invariant magnetic flux enclosed by a poloidal field line, and the field-line angular velocity $\Omega_F = F_{tr}/F_{r\phi} = F_{t\theta}/F_{\theta\phi}$. 
$\Omega_F$ does not correspond to a physical rotation of the system \footnote{In particular, field-line rotation generically gives rise to a light cylinder surrounding the black hole, beyond which the four-velocity of a corotating observer becomes spacelike.}. Instead, it determines the structure of the induced electric field $E^{\mu}$ and regulates the efficiency of energy extraction in the black hole magnetosphere \cite{Blandford:1977ds} together with frame dragging. 
Using $\psi$ and $\Omega_F$, the pseudo-magnetic field can be expressed as
\begin{equation}
	B^r=\frac{\partial_\theta \psi}{\sqrt{-g}},\,\,\,\, B^\theta=-\frac{\partial_r \psi}{\sqrt{-g}},\,\,\,\, B^\phi=(u^\phi-\Omega_Fu^t)\frac{B^r}{u^r}.\label{Bform}
\end{equation}
where $g=\det(g_{\mu\nu})$. 
When evaluating the synchrotron emission profile of a magnetized disk, it is convenient to employ the covariant magnetic field, $b^{\mu} = -(*F)^{\mu\nu}u_{\nu}$, measured in FCF.
Under the ideal-MHD condition, $b^{\mu}$ is related to the pseudo-magnetic field through $b^t = u_{\mu}B^{\mu}$, $u^t b^i = B^i + b^t u^i$. Using Eq.~\eqref{Bform}, we obtain
\begin{equation}
	b^\mu=-\frac{\partial_\theta \psi}{\sqrt{-g}u^r}\left[(u_t+\Omega_Fu_\phi)u^\mu+\delta_t^\mu+\Omega_F\delta_\phi^\mu\right] \,.\label{bform}
\end{equation}
As the horizon is approached, the components $b^t, b^{\phi}$ diverge owing to the use of BL coordinates. However, physically measurable quantities---such as the field magnitude $b^2 = b^{\mu}b_{\mu}$---remain finite throughout the entire black-hole exterior (as can be seen from Fig.~\ref{fig:bwithr}).
For a flow following conical streamlines, $\psi$ only depends on $\theta$. 
In this work we adopt a simple split-monopole configuration \cite{Blandford:1977ds}, $\partial_\theta \psi = -\mathrm{sign}(\cos\theta)\sin\theta$, in the disk region. 

Regarding the field-line angular velocity $\Omega_F$, numerical simulations indicate that magnetized disks typically satisfy $\Omega_F \sim 0.2\,\Omega_H$ to $0.5\,\Omega_H$ \cite{Tchekhovskoy:2011zx, McKinney:2012vh}. Note that the ideal-MHD description is not only well suited for magnetized jets \cite{Song:2025mhj}, but could also apply to the main emission region of the accretion flow, since the dominant synchrotron radiation arises from disk zones with smaller plasma beta $\beta$ \cite{Davelaar:2023dhl}.

Fig.~\ref{Bmuplot} depicts the magnetic field in the equatorial plane for several representative parameter sets \footnote{When referring to “within the equatorial plane,” we in fact mean a plane arbitrarily close to but not coincident with the equatorial plane, since the magnetic field vanishes exactly at $\theta = \pi/2$ in the split-monopole solution. And we have chosen $\theta=\pi/2-0.01$ in Fig.~\ref{Bmuplot}.}. 
As the flow approaches the horizon, a free-falling stream with zero angular momentum is forced to co-rotate with the spacetime, attaining an angular velocity $\Omega_\text{drag} = \omega \approx 2a r^{-3}$. 
Once the inflow enters the ergoregion (the gray areas), it must co-rotate with the spacetime regardless of its prior motion. The quantity $\Omega_\text{drag}$ serves as a characteristic indicator of the magnetic‑field geometry. We can therefore express $B^\phi$ as $B^\phi \approx B^r  \left(\Omega_\text{drag}- \Omega_{F}\right)\alpha^{-2}$, where $\alpha$ is the lapse function, which vanishes as $r \to r_+$. Thus, $B^{\mu}$ becomes predominantly toroidal at the horizon, where the field line winding ($\eta = |rB^{\phi}/B^r|$) goes to infinity. 

As illustrated in Fig.~\ref{Bmuplot}, for a non-rotating field line ($\Omega_F = 0$), the field configuration transitions from a radial to a toroidal structure near the horizon. When $\Omega_F \neq 0$, the field pattern switches from counter-clockwise to clockwise as the radius decreases, with a critical transition occurring at $r_\text{c}\approx \left( 2a \Omega_{F}^{-1}\right)^{1/3}$, denoted by  dashed blue curves.
As the field lines bend near $r = r_\text{c}$, it may serve as a potential site for magnetic reconnection. In our model, this bending is not very pronounced because the toroidal velocity remains moderate; however, for other flow patterns, the effect could be much stronger and therefore merits further investigation. In addition, since we adopt a split-monopole magnetic configuration, the most idealized location for magnetic reconnection is the near-equatorial region, where the magnetic field sharply reverses its direction.

Similar to the flow density and temperature, the magnetic field strength shows only a weak dependence on the black hole parameter (as illustrated in Fig.~\ref{fig:bwithr}, Appendix.~\ref{appplot}). In the case of $a = 0.5$, the profiles of $b$ for $h = 0$ and $h = -0.9$ nearly coincide, while for $a = 0.9$, changing $h$ from $0$ to the relatively small value of $-0.27$ results in variations of $b$ of less than $1\%$. 

In addition, the variation of $\Omega_F$ is found to have a negligible effect on $b$ in our case.  
We further note that the field-line rotation and winding exert only a minor influence on the observed intensity distribution.  
However, they are expected to leave discernible imprints on the synchrotron polarization pattern, provided a correctable Faraday effect.  
In the FCF, the emitted linear polarization vector is perpendicular to the local magnetic field, and its subsequent evolution is determined by parallel transport in the curved spacetime \cite{Zhang:2023cuw, Chen:2024jkm}.  
Detailed analysis of the polarization structures will be presented in future work.

\section{Observational Features}
\label{sec4}

\subsection{Imaging scheme}

To evaluate the variation of light intensity along null geodesics, we employ the General Relativistic Radiative Transfer (GRRT) equation \cite{Lindquist:1966igj}. In covariant form, it can be written as an evolution equation along a null geodesic parameterized by the affine parameter $\lambda$: 
\begin{equation}
	\label{grrt}
	\frac{\dd}{\dd\lambda}\left(\frac{I_\nu}{\nu^3}\right)=\frac{j_\nu-\alpha_\nu I_\nu}{\nu^2}\,.
\end{equation}
Here, $\nu$ is the emission frequency measured in the FCF, $I_\nu$ is the specific intensity, and $j_\nu$ and $\alpha_\nu$ are the emissivity and absorption coefficient in FCF, respectively. 
Note that $\nu$ varies along the geodesic for fixed observational frequency $\nu_o$, as a result of gravitational and Doppler shifts.
After getting the photon trajectories via ray-tracing the geodesic equation Eq. \eqref{eom}, the intensity distribution $I_\nu$ at each point on the observer's screen is obtained by integrating the transfer equation \eqref{grrt} along these traced-back trajectories, under a specific emission model. 

For the synchrotron emission by thermal electron distribution within a fluid volume, the peak frequency of the spectrum is located around $\nu_\text{peak} = 5\Theta_\text{e}^2\nu_B$, where $\nu_B = eb/(2\pi m_\text{e})$ is the cyclotron frequency \cite{1979Lightman}. 
The emissivity of thermal synchrotron radiation can be calculated analytically in the asymptotic regimes $\nu \ll \nu_\text{peak}$ and $\nu \gg \nu_\text{peak}$. For the intermediate regime, a fitting formula that well describes the behavior of $j_\nu$ across varying model parameters is given by \cite{leung2011numerical}
\begin{equation}
	\label{jnu}
	\begin{aligned}
		j_\nu=\frac{2^{-1/2}\pi\text{e}^2}{3}\frac{n_\text{e}\nu}{\Theta_\text{e}^2}\left(1+2^{11/12}X^{-1/3}\right)^2\exp(-X^{1/3}) \,, 
	\end{aligned}
\end{equation}
where $X$ measures the ratio between the emission frequency and the peak frequency:
\bea
X =  \frac{45\nu}{2\nu_\text{peak}\sin{\theta_B}} \,,
\eea
with $\theta_B = \arccos{\left(b^{-1}b^{\mu}k_{\mu}\right)}$ the pitch angle between the magnetic field and emission direction within the FCF. 
Finally, as the photon energy in the millmeter band is much smaller than the electron thermal energy (even considering gravitational redshift), the Rayleigh-Jeans approximation applies. The self-absorption coefficient of thermal electrons is therefore given by $\alpha_\nu = j_\nu/(2\nu^2k_\mathrm{B}T_\text{e})$.  At 230 GHz, the absorption is very weak and becomes appreciable only for rays with enough long optical paths near the photon sphere (the near-critical rays); nevertheless, it is fully accounted for in the subsequent calculations.


\subsection{Intensity maps}

\begin{figure*}[htbp]
	\centering
	\includegraphics[width=6.4in]{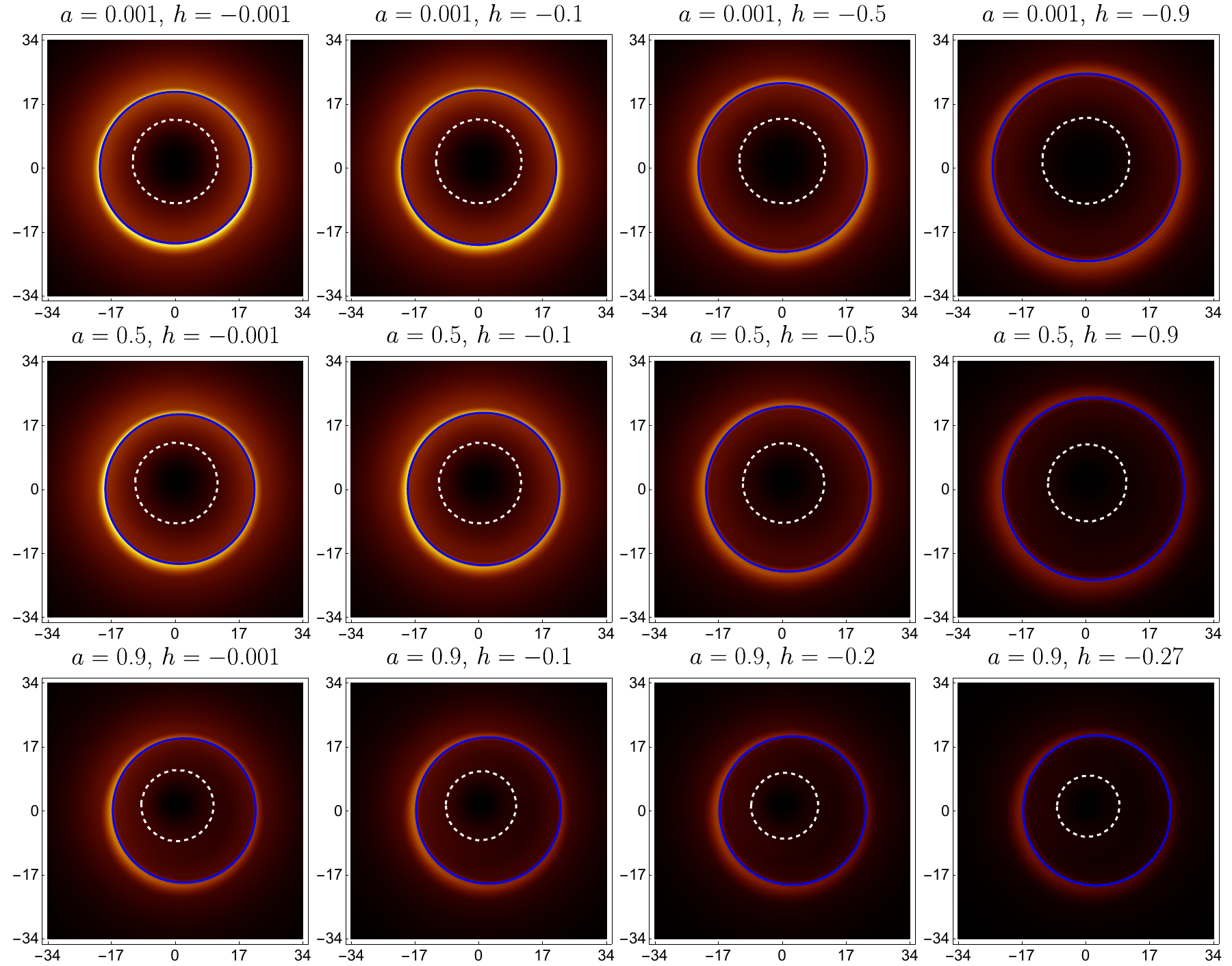}\\
\quad \quad \,\,\, \includegraphics[width=6.2in]{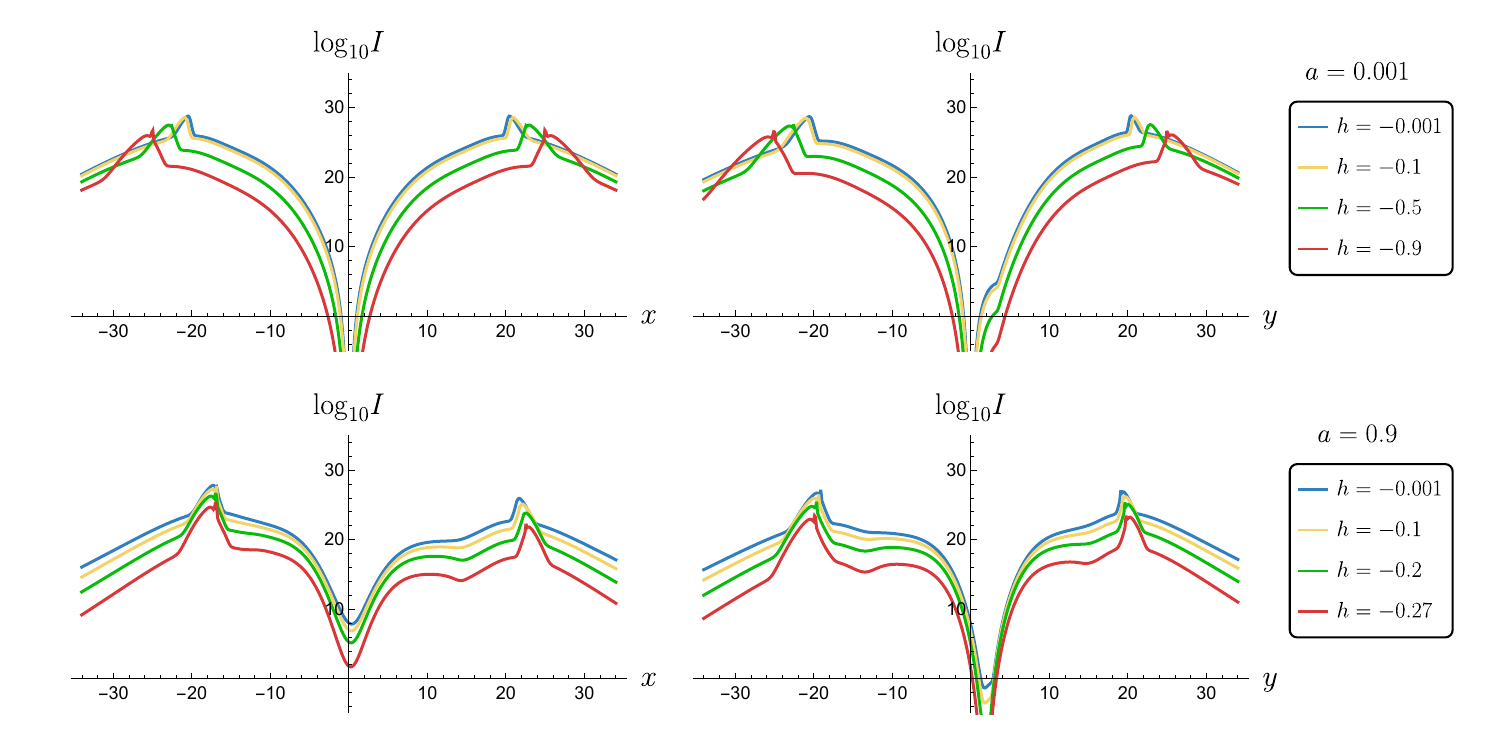}
	\caption{Imaging results of the thick accretion disk with a free-falling flow ($E=\mu$) around a rotating Horndeski black hole viewed at an inclination angle of $\theta_o=17^\circ$. 
The solid blue curve marks the critical curve, traced by photons orbiting the black hole infinitely many times. The dashed white curve denotes the inner shadow boundary, corresponding to the direct image of the event horizon \cite{Chael:2021rjo}. 
The disk parameters are $z = 20, \sigma=0.2$ and $\Omega_F=0.2\,\Omega_H$.
The bottom two rows display the intensity profiles along the $x$-axis and $y$-axis for different black hole parameters.}
	\label{fig17}
\end{figure*}

We model the transport of disk emission to a distant ZAMO using GRRT \cite{Zhang:2024lsf}, and produce the 230 GHz images of rotating hairy black holes in Horndeski theory.
Fig.~\ref{fig17} shows the intensity maps on the observer's screen at an inclination angle of $\theta_o=17^\circ$, for various spin and hair parameters \footnote{As shown in the Fig.~\ref{parah}, the allowed range of the parameter $h$ for rotating black hole solutions depends on $a$. In numerical calculations we set $h\in[0,-0.9]$ for $a=0.001$ and $a=0.5$, while for the case $a=0.9$ we adopt $h\in [0, -0.27]$.}. For clarity, we also display the corresponding one-dimensional intensity cuts along the horizontal and vertical directions.
For comparison, the results at a higher inclination angle are presented in Fig.~\ref{fig80} in the Appendix.~\ref{appplot}.

The accretion inflow is assumed to be free-fall with $E=\mu$. The disk thickness parameter is fixed at $\sigma=0.2$, and the ion-to-electron temperature ratio is set to $z =20$.  
The magnetic-field strength at the event horizon is normalized to 10 G, consistent with the estimates for M87*. We further adopt a conservative choice for the field-line angular velocity, $\Omega_F = 0.2\,\Omega_H$; testing $\Omega_F = 0$ yields only marginal differences in the resulting images.

At 230 GHz, thermal electron self-absorption is weak, so the plasma is largely optically thin and a clear ring structure appears in all images. This ring originates from photons that orbit the black hole multiple times before escaping to the observer \cite{Gralla:2019xty}.
In addition to the low optical depth, lensing-induced frequency shift, combined with pitch-angle effect, further amplifies the intensity near the ring by 3-5 orders of magnitude, making the ring feature particularly prominent in our disk model \cite{Zhang:2024lsf}.
It can be seen that as the hair parameter decreases, the ring size systematically increases. For instance, in the nearly non-spinning case, reducing $h$ from 0 to $-0.9$ enlarges the ring diameter from 40.37 $\mu$as to 49.90 $\mu$as.
Moreover, the image brightness also shows a significant dependence on the hair parameter. 
As $h$ decreases, the intensity gradually weakens. 

Close to the horizon, the strong gravitational redshift suppresses the observed intensity and produces a central shadow region in the images, as shown in Fig.~\ref{fig17}. In each plot, we also outline the contour of the projected horizon onto the screen, which is also referred to as the ``inner shadow''. 
Such a theoretically predicted quantity is shaped by gravity and the inclination angle, making it a potential probe of the spacetime geometry in addition to the photon ring \cite{Chael:2021rjo}. 
As the parameter $h$ decreases, the size of the inner shadow gradually decreases, consistent with the trend of a shrinking event horizon radius \cite{Walia:2021emv, Afrin:2021wlj}. However, radiation from outside the equatorial plane in the thick disk blurs the inner shadow boundary, making it difficult to distinguish, unlike in the thin disk case \cite{Hou:2022eev}. 

A vertical comparison of the first two columns in Fig.~\ref{fig17} reveals that an increase in the spin parameter $a$ not only reduces the size of the central shadow but also introduces a top-bottom asymmetry. 
This effect is more evident in the vertical cuts presented in the same figure.
However, compared with the effect of $h$, the influence of $a$ on the diameter of the photon ring is relatively minor.

Appendix.~\ref{appplot} includes the results with $\theta_o=80^\circ$.  
Similar to the $\theta_o=17^\circ$ case, as $h$ decreases, the photon ring expands, while it brightness decreases.  
In Fig.~\ref{fig80}, it is evident that the peaks in the horizontal intensity cut gradually shift to the right as $a$ increases or $h$ decreases, with the left peak showing a more pronounced shift.  
This results in the ring deforming into a ``D-shape'', consistent with previous studies.
Furthermore, the intensity in the upper shadow region ($y>0$) is on average about two orders of magnitude lower than that in the lower shadow region ($y<0$).

\subsection{Brightness suppression}

\begin{figure*}[htbp]
	\centering
	\includegraphics[width=6.2in]{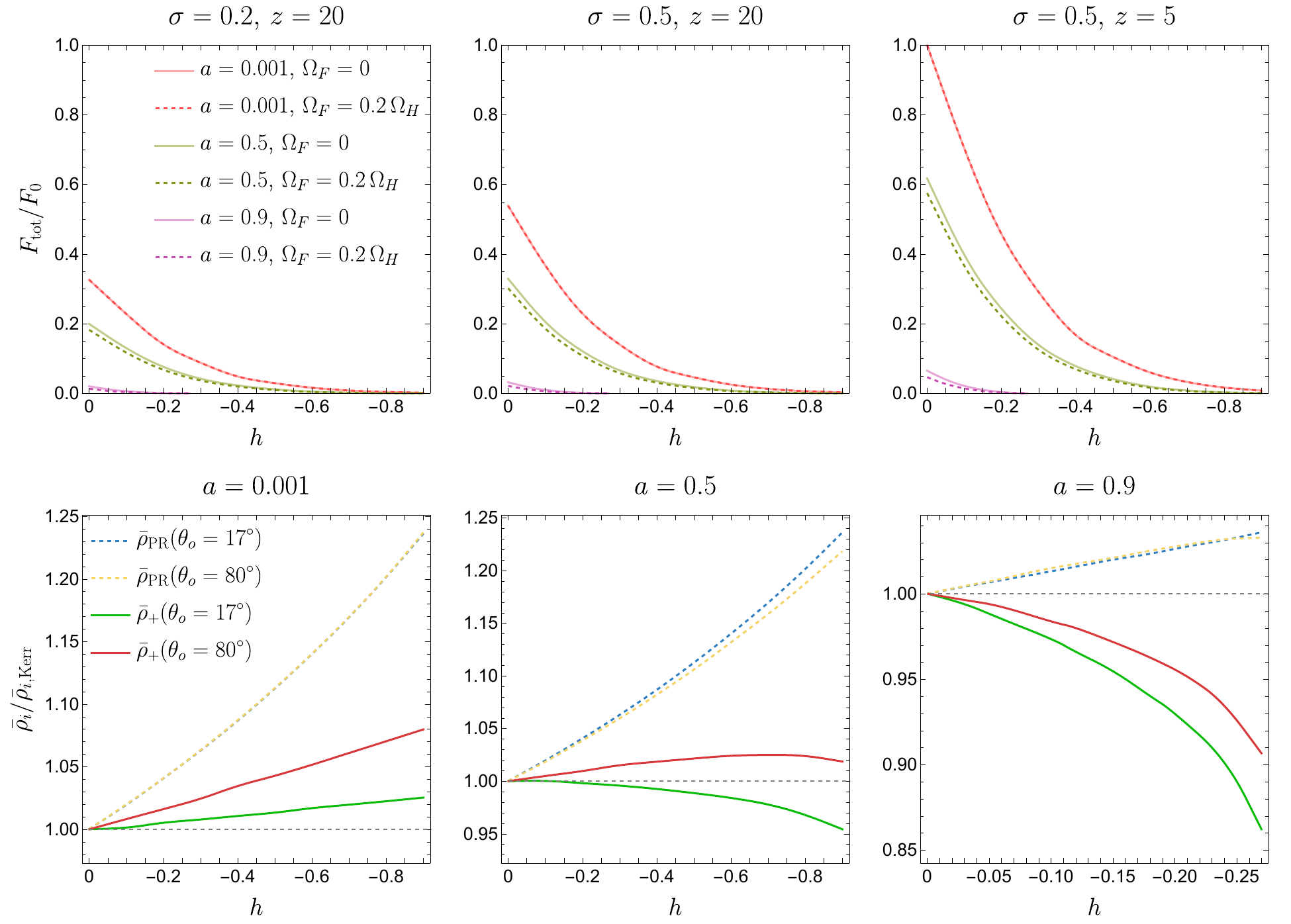}
	\caption{Variation of typical observables in black hole image with the hair parameter $h$. The top row of panels depicts the variation of the normalized total flux $F_\text{tot}/F_0$ with parameter $h$ for different parameters at an observational inclination angle of $\theta_o=17^\circ$, where $F_0$ is the total flux with $a=0.001$, $h=0$, $\sigma=0.5$, $z=5$ and $\Omega_F=0$. The bottom row of panels characterizes the variation of the rescaled average radius $\bar{\rho}_i/\bar{\rho}_{i,\text{Kerr}}$ with $h$, where the dashed and solid lines represent the results for the photon ring (critical curve) and the inner shadow boundary curve, respectively, and the grey dashed line indicates the result for a Kerr black hole.}
	\label{figfr}
\end{figure*}

To gain deeper insight into how the hair parameter affects the photon ring, the inner shadow, and the overall image brightness, we perform a quantitative analysis of representative imaging features.  
In the top row of Fig.~\ref{figfr}, we show how the $17^\circ$ total flux $F_\text{tot}$ varies with $h$ for different spin and disk parameters, namely the ion-to-electron temperature ratio $z$, the thickness parameter $\sigma$, and the field-line angular velocity $\Omega_F$. For ease of comparison, the total flux is normalized to $F_0$, corresponding to the parameter set $a=0.001$, $h=0$, $\sigma=0.5$, $z=5$, and $\Omega_F=0$.

It is observed that increasing $a$ or decreasing $h$ leads to a pronounced reduction in the total flux.  
As discussed earlier, the parameter $h$ exerts only a minor influence on the flow structure (see Fig.~\ref{fignT}); therefore, the decrease in total flux is primarily attributed to gravitational redshift. 
To illustrate this, we consider a simplified yet instructive scenario that the emission is from the equatorial plane, viewed face-on. 
In this case, the redshift factor is approximated as
\begin{equation}
	\label{redshift}
	g  = \f{\nu_o}{\nu} \approx \left\{
	\begin{aligned}
		& \,\, 1-\sigma_r \sqrt{2|\Phi|} \qquad \text{(free-falling flow)} \\
		&\,\,  E^{-1} \left(1 - 2|\Phi|\right)  \quad  \text{(circular flow)}
	\end{aligned}\right.
\end{equation}
where $\nu = -k^{\m}u_{\m}$ denotes the emission frequency, and $\Phi$ is the effective potential defined in Eq.~\eqref{Phieff}. For the free-falling case, when the photon is emitted outward with a positive radial direction indicator $\sigma_r = 1$, $g$ decreases as the effective potential increases. In contrast, for inward emission $\sigma_r = -1$, gravity induces a blue-shift effect ($g > 1$) because the inflow velocity is enhanced.

A decrease in $h$ signifies a stronger accumulation of positive energy scalar fields within the main emission region, which enhances $|\Phi|$ and consequently leads to a stronger gravitational redshift for outward emission ($\sigma_r = 1$).
To clarify how the variation of $h$ affects the observed intensity, we examine the change in the observed intensity produced by a small perturbation in $h$:
\begin{equation}
	\label{grrt1}
	\d I_{\n_o} \approx \, \nu_o \,\d h \int_{\lambda}\dd\lambda \,\f{\pl g}{\pl h} \left( 2 g j_{\nu}  - \nu_o \f{\pl j_{\nu}}{\pl \n}\,\right) \,,
\end{equation}solution for $I_{\n_o}$ obtained by integrating Eq.~\eqref{grrt} under the assumption of negligible absorption \footnote{We have neglected the change in photon trajectories resulting from variations in $h$, as this effect is expected to be subdominant compared to the redshift.}.
The first term on the r.h.s. of Eq.~\eqref{grrt1} originates from the relativistic beaming effect intrinsic to radiative transfer \cite{Lindquist:1966igj}, 
while the second term accounts for the frequency shift between the emitted and observed photons. 

For outward emission which accounts for the majority of photons contributing to the total flux, we have $\pl g/\pl h > 0$ for negative $h$. 
Meanwhile, as the 230\,GHz observing band lies on the high-frequency side of the peak of the thermal synchrotron spectrum of the accreting electrons, $\nu_o > \nu_{\mathrm{peak}}$, the spectral slope of $j_{\n}$ is negative, with $j \propto \n\, e^{-\n^{1/3}}$.  
A decrease in $g$ therefore corresponds to a higher intrinsic emission frequency and a lower emissivity (spectral-shift effect).
As a result, the terms on the r.h.s. of Eq.~\eqref{grrt1} have the same sign and jointly reduce the observed brightness for $\d h < 0$.

For the strongly lensed rays that constitute the narrow ring, a large fraction of photons are emitted outside the photon sphere, moving inward across the sphere before eventually escaping to infinity. These photons have a negative radial direction indicator ($\sigma_r = -1$) and experience a blue-shift $g > 1$. 
Consequently, they enhance the ring brightness through relativistic beaming and frequency-shift, which accounts for the ring being much brighter than the direct emission in the disk model. 
Although photons emitted from the inner side of the photon sphere still have $g<1$ and contribute a suppressed intensity, due to the outer photon sphere's contribution, the overall diminution of the ring brightness with increasing $|h|$ is smaller than the corresponding decline in the direct-image brightness, as seen in the horizontal and vertical intensity profiles in Fig.~\ref{fig17} and Fig.~\ref{fig80}.

In addition, disk parameters also influence the 230\,GHz flux.  
Comparison among different panels shows that a larger disk thickness, or a smaller ion-to-electron temperature ratio $z$, tends to increase the observed total flux.  
The effect of the field-line angular velocity $\Omega_F$ is comparatively minor---increasing $\Omega_F$ slightly reduces the total flux, particularly in high-spin cases.

\subsection{Ring and shadow deformation}

The theoretical shapes of the photon ring and inner shadow directly encode the underlying spacetime geometry, although they remain difficult to observe. It is nevertheless useful to examine how these features depend on the hair parameter.

For a closed convex curve, its geometric centers on the image plane are defined as $x_c=(x_\text{max}+x_\text{min})/2$ and $y_c=(y_\text{max}+y_\text{min})/2$, where the subscripts “max” and “min” denote the maximum and minimum values of the horizontal and vertical coordinates, respectively.  
Introducing a polar coordinate $(\rho,\alpha)$ centered at $(x_c,y_c)$, the average radius of the curve is \cite{Hou:2022eev}
\begin{equation}
	\bar{\rho}_i=\frac{1}{2\pi}\int_0^{2\pi}\rho_i(\alpha)\dd\alpha\,.
\end{equation}
The bottom row of Fig.~\ref{figfr} displays the variations of the average radii of the photon ring and the inner shadow boundary as functions of $h$.  
For convenient comparison, the vertical axis is expressed in terms of the normalized average radius $\bar{\rho}_i/\bar{\rho}_{i,\text{Kerr}}$, which effectively characterizes the deviation of each curve from its Kerr counterpart.  

The average radii of both the photon ring and the inner shadow boundary deviate from the Kerr case to different extents as $h$ decreases.  
Specifically, the photon ring shows a stronger sensitivity to variations in $h$ at moderate spins ($a = 0.001, 0.5$), whereas at higher spin ($a = 0.9$), the inner shadow boundary becomes more responsive to $h$.  
Notably, for $a=0.5$,\,$\bar{\rho}_{+}$ first increases and then decreases with decreasing $h$, suggesting the existence of a degeneracy between $a$ and $h$ in certain parameter ranges.  

Moreover, the deviation of the photon ring from its Kerr counterpart is found to be largely insensitive to the inclination angle, particularly at low spins, in agreement with \cite{Johannsen:2010ru}.
Regarding the inner shadow boundary, the radius $\bar{\rho}_+$ gradually decreases with increasing spin and eventually becomes smaller than $\bar{\rho}_{+,\text{Kerr}}$.  
This behavior arises from the progressive deformation of the inner shadow as the spin increases.

\section{The First Photon Ring}
\label{sec5}

\begin{figure*}[htbp]
	\centering
	\includegraphics[width=5in]{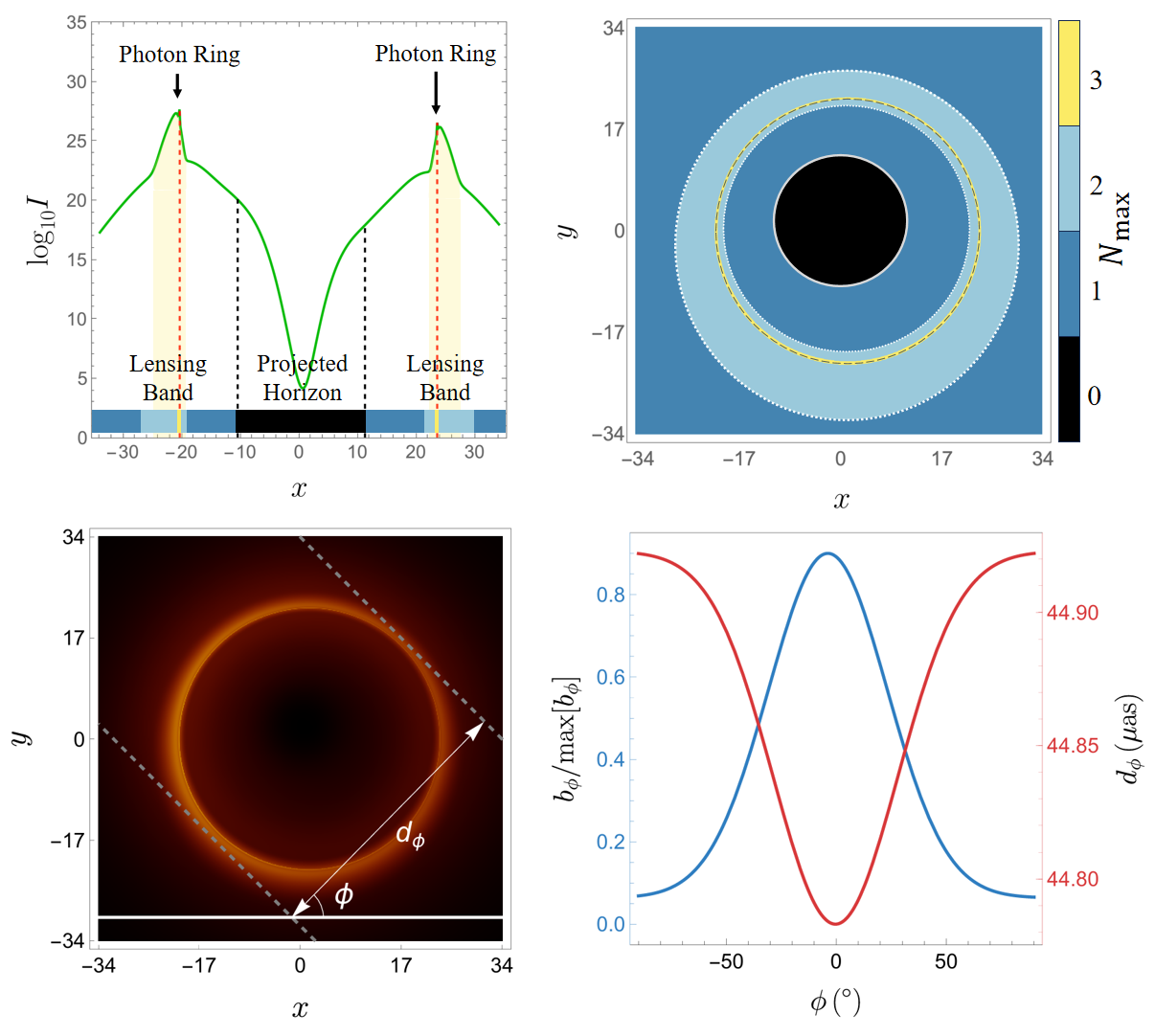}
	\caption{Schematic diagram of black hole image and typical observables. \textbf{(Top Left):} Illustration of imaging features in an intensity cut. \textbf{(Top Right):} Maximum number of equatorial crossings. $N_{\text{max}} = 0, 1, 2, \ldots$ correspond to the projected horizon (inner shadow), direct emission, lensing band, etc \cite{Gralla:2019xty}. \textbf{(Bottom Left):} Schematic of the FPR and the diameter $d_\phi$. \textbf{(Bottom Right):} Typical behavior of $d_{\phi}$ and $b_{\phi}$ as functions of the polar angle. In all plots, the black hole parameters are $a = 0.5$, $h=-0.5$; the thick disk's parameters are $E = \mu$, $z =20$, $\sigma=0.2$ and $\Omega_F=0.2\,\Omega_H$; the inclination angle is $\theta_o = 17^{\circ}$.}
	\label{figtimes}
\end{figure*}

The BHEX mission aims to extend VLBI into space by deploying a satellite in Earth orbit at an altitude of $\gtrsim 20000$km, thereby achieving unprecedented angular resolution and enabling the sharpest black hole imaging to date \cite{Johnson:2024ttr}. 
The Telescope's angular resolution is given by 
\begin{equation}
	\theta_\text{res} \approx 10 \,\mu\text{as} \, \left(\f{\lambda}{1 \text{mm}}\right) \left(\f{b}{2 \times 10^4 \text{km}}\right)^{-1} \,,
\end{equation}
where $\lambda$ is the observed wavelength and $b$ is the baseline length. A space baseline of $b\approx 30,000$ km yields an angular resolution of $\theta_\text{res}\approx 8.7 \,\mu$as at 230 GHz and $\theta_\text{res}\approx 5.8\,\mu$as at 345 GHz. Thus, BHEX has the potential to resolve the ``first photon ring'' (FPR), which may encode key signatures of strong-gravity physics around black holes.

The FPR is defined as the intensity distribution within the ``lensing band'', generated by photons completing one additional orbit around the black hole, corresponding to a maximum number of equatorial crossings $N_{\text{max}} = 2$ \cite{Gralla:2019xty}, as illustrated in the top row of Fig.~\ref{figtimes}. 
In the case of a geometrically and optically thin disk, the image 
can be cleanly decomposed into multiple sub-images labeled by $N_{\text{max}} = 0, 1,2,\ldots$  
In more realistic disks, 
when absorption is weak and the disk thickness remains well-defined---as in our case---$N_{\text{max}}$ still serve as a reliable descriptor for organizing images \cite{Vincent:2022fwj}. 

The FPR diameter along the $\phi$-direction on the image plane, denoted $d_{\phi}$, is defined as the separation between the brightness peaks in the two corresponding quadrants, as shown in the bottom-left panel of Fig.~\ref{figtimes}. 
In the visibility domain, $d_{\phi}$ determines the associated interferometric diameter, which can be extracted by fitting the visibility amplitude over the range of baselines where the FPR contribution dominates \cite{Paugnat:2022qzy}
\begin{eqnarray}
	\begin{aligned}\label{vis}
&|V(u,\phi)|\approx \sqrt{(A_\phi^L)^2+(A_\phi^R)^2+2A_\phi^LA_\phi^R\sin(2\pi d_\phi  u)}\,, \\
&A_\phi^{L/R}(u)= \f{1}{2} \left[\,e_\text{up}(u)\pm e_\text{low}(u)\,\right] \,,
	\end{aligned}
\end{eqnarray}
where $e_\text{up}(u)$, $e_\text{low}(u)$ denote the upper and lower envelopes of the oscillatory pattern in Eq.~\eqref{vis}, whose periodicity is set by $1/d_\phi$. 
In addition, the FPR's brightness distribution can likewise be inferred from the visibility by integrating $|V(u,\phi)|^2$ over the baseline interval where the FPR dominates \cite{Farah2025}: $b_\phi\equiv\int_{u_-}^{u_+}|V(u,\phi)|^2\dd u$.
The angular profiles of $d_\phi$ and $b_\phi$ can be well described by Gaussian- or sinusoid-like functions, as shown in the bottom-right panel of Fig.~\ref{figtimes}.

The maximal interferometric diameter of the FPR is given by the maximum of $d_\phi$, denoted by $d_+$, which is sensitive to the black hole’s spin, whereas the full width at half maximum of $b_\phi$ primarily traces the viewing inclination \cite{Farah2025}. 
However, the FPR generally overlaps with both the direct emission and the higher‑order photon rings in the visibility domain, and this interference becomes increasingly significant at higher inclination angles \cite{Cardenas2023}. 
At the same time, the ability of $d_+$ to constrain the spin deteriorates markedly at large inclinations \cite{Farah2025}. Therefore, our discussion of the FPR's interferometric characteristics is restricted to the case $\theta_o=17^\circ$.

\begin{figure*}[htbp]
	\centering
	\includegraphics[width=6.2in]{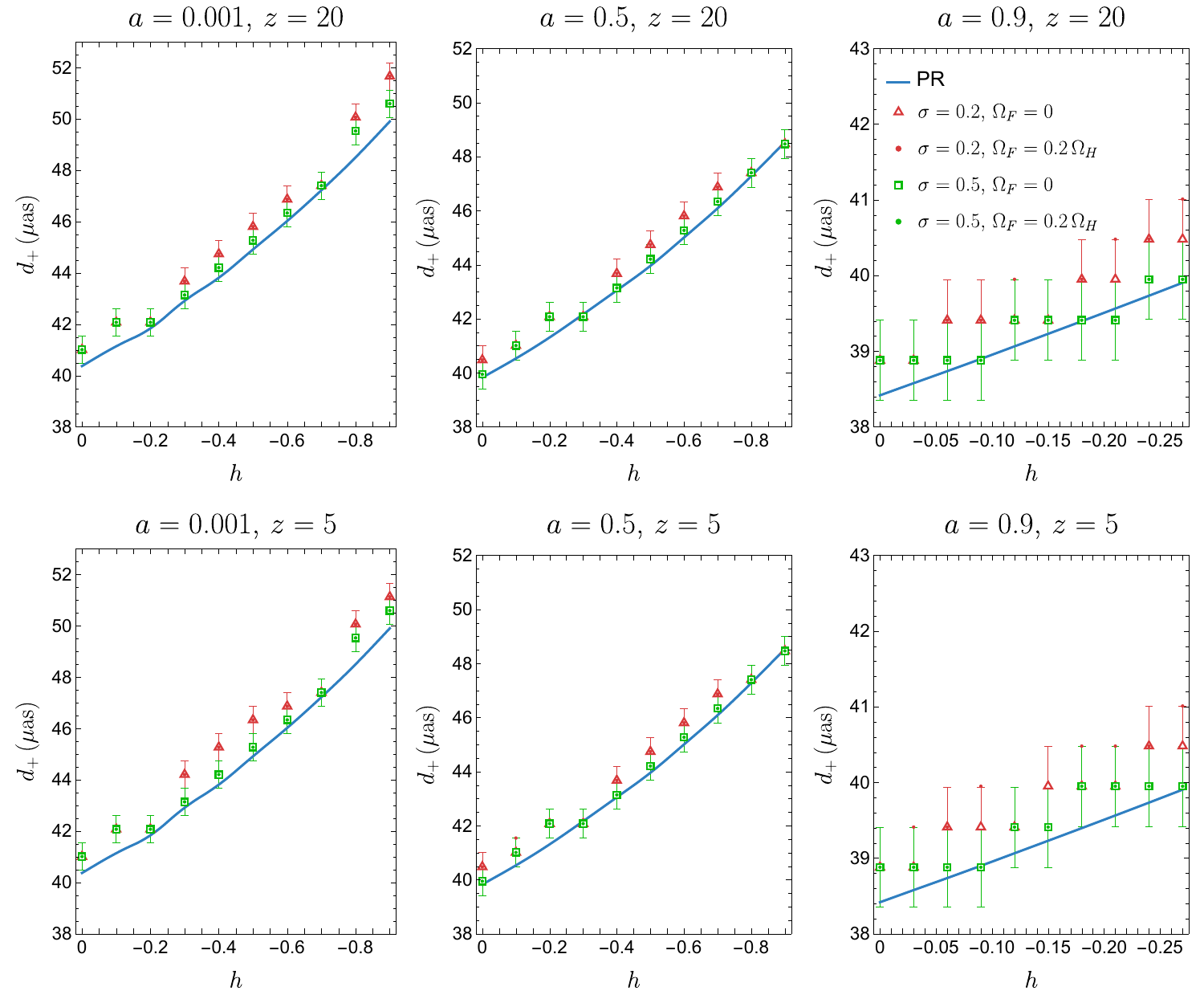}
	\caption{Variation of the maximum interferometric diameter $d_+$ with scalar hair parameter $h$ under different accretion disk parameters. From top to bottom, the rows correspond to $z=20$ and $z=5$, respectively. From left to right, the columns correspond to $a=0.001$, $0.5$, and $0.9$, respectively. The observational inclination angle is fixed at $\theta_o=17^\circ$. Error bars of $\pm 0.5 \mu$as represent the numerical accuracy. The solid blue lines indicate the diameter of the photon ring evaluated at the same polar angle direction as the FPR.}
	\label{dmax}
\end{figure*}

We now evaluate the feasibility of using $d_+$ to probe the black hole’s scalar hair in Horndeski theory. 
Unlike the extremely narrow photon ring, the FPR spans a finite-width lensing band whose detailed intensity structure depends on the accretion flow's emission properties.
To evaluate the robustness of $d_+$  against uncertainties in the flow, we compute its value across a range of accretion-model parameters and compare the outcomes.
Fig.~\ref{dmax} illustrates how $d_+$ varies with the scalar-hair parameter $h$ for different choices of the temperature ratio $z$, disk-thickness parameters $\sigma$ and field-line angular velocities $\Omega_F$.
The numerical precision of our calculations is set to $1 \,\mu$as, corresponding to the maximum resolution achievable by space-based VLBI missions \cite{Johnson:2019ljv}. Accordingly, only variations in physical parameters that induce changes in $d_+$ exceeding $1 \,\mu$as can be considered observationally significant.

On average, the results show that $d_+$ is about 3\% larger than the diameter of the photon ring (critical curve) evaluated at the same $\phi$-direction (solid blue lines), and exhibits the same qualitative dependence on $h$. 
Increasing the disk thickness further narrows the gap between $d_+$ and the photon-ring diameter.
Moreover, the values of $d_+$ obtained for different field-line angular velocities $\Omega_F$ and temperature ratios $z$ nearly coincide; minor deviations appear only at a few data points for $a=0.9$ (three points for $z=20$ and five for $z=5$). 
This arises because, at high spin, the black hole's angular velocity $\Omega_H$ becomes large, leading to more pronounced changes in the magnetic-field configuration between $\Omega_F = 0$ and $0.2\,\Omega_H$, as illustrated in Fig.~\ref{Bmuplot}. 
These structural differences modify the emission profile and therefore the resulting images.
Overall, however, $d_+$ remains largely insensitive to detailed variations in the accretion-flow and magnetic-field properties.

Variations in the spacetime parameters, either $a$ or $h$, induce corresponding changes in $d_+$. 
As shown in Fig.~\ref{dmax}, their effects differ qualitatively: increasing the spin parameter decreases $d_+$, while decreasing $h$ enlarges it. The impact of $h$ is also substantially stronger. For example, in the case of $z=20$ (top row of Fig.~\ref{dmax}),
\begin{itemize}[leftmargin=0pt]
	\item Fixing $h=0$ and increasing $a$ from 0.001 to 0.9 reduces $d_+$ only mildly, from $41.01\,\mu$as to $38.88\,\mu$as. This weak dependence reflects the fact that, for Kerr black holes, the photon-ring diameters vary only mildly with spin \cite{Johannsen:2010ru}.
	\item In contrast, fixing $a=0.001$ and $\sigma=0.2$ while decreasing $h$ from 0 to $-0.9$ leads $d_+$ to rise markedly from $41.01\,\mu$as to $51.66\,\mu$as. Even in the high-spin case ($a=0.9$), where the allowed range of $h$ is limited to $[-0.2721,0]$, the resulting variation in $d_+$ remains as large as $1.61\,\mu$as (from $40.49\,\mu$as to $38.88\,\mu$as), well within BHEX's detection capability.
\end{itemize}
Therefore, for small inclination angles, $d_+$ can serve as an effective observable for constraining the hair parameter $h$, at least enabling a clear distinction between large-$h$ from small-$h$ configurations for moderately spinning black holes.


\section{Summary}
\label{sec6}

In this work, we constructed an analytical horizon-scale accretion disk model around a hairy rotating black hole in Horndeski theory and examined the resulting 230 GHz images. We focused on how the scalar hair parameter modifies the flow structure, total flux, intensity distribution, and lensed rings, and in particular demonstrated that the maximal interferometric diameter of the fisrt photon ring provides a promising probe of scalar hair.

We found that the hair induces only mild changes in the inflow and magnetic-field structures that set the intrinsic emissivity. However, it strengthens the gravitational redshift, substantially dimming both the direct image and the total flux through relativistic transfer effects and a shift in the emissivity spectrum. By contrast, many photons forming the lensed and photon rings originate from inward emission and experience net blueshift. As a result, although the total flux decreases, the photon ring is attenuated less than the direct component, making the ring comparatively more pronounced.
In addition, the theoretical critical curve and inner shadow respond differently to variations in hair, spin, and inclination. Notably, the critical curve is more sensitive to $h$ at moderate spins, while the inner shadow becomes more sensitive to $h$ at high spins.

We place particular emphasis on assessing the ability of the fisrt photon ring's maximal interferometric diameter $d_+$ to probe the black hole hair parameter. We domonstrated that for a viewing angle of $\theta_o = 17^{\circ}$, $d_+$ is largely insensitive to the details of the accretion flow and magnetic-field structure. 
Moreover, its response to variations in the hair parameter $h$ is significantly stronger than its dependence on the spin parameter or any other accretion-flow parameter. 
We therefore conclude that, within BHEX resolution and under modest inclination, the interferometric diameter of the fisrt photon ring can serve as a robust observable for distinguishing large-$h$ from small-$h$ configurations. 
Looking forward, combining $d_+$ with additional image-based observables—such as the resolved polarimetric structure across the image plane and the azimuthal brightness profile of the first photon ring—may further strengthen constraints on the scalar hair.

Moreover, since scalar hair can influence both the dynamics of the accretion flow and the redshift distribution of its emitted radiation, multiwavelength observations offer a complementary avenue for investigation. In particular, broadened iron lines observed in X-ray reflection spectra of both stellar-mass and supermassive black hole candidates are expected to be highly sensitive to the underlying spacetime \cite{Miller:2007tj, Cackett:2009ih, Dauser:2010ne, Bambi:2012at, Fabian:2014tda, Johannsen:2014arp, Bambi:2016sac}, and may therefore provide independent probes of black hole hair. We will explore these possibilities in future work.

\section*{acknowledgements}

We thank Peng-Cheng Li and Zhenyu Zhang for their assistance. The work is partly supported by NSFC Grant No. 12575048, 12205013 and 12547123. M. G. is also supported by Open Fund of Key Laboratory of Multiscale Spin Physics (Ministry of Education), Beijing Normal University and he is also supported by the BNU Tang Scholar.

\bibliographystyle{utphys}

\bibliography{Horndeski_BAAF}

\clearpage

\appendix

\onecolumngrid

\section{Supplementary Plots}
\label{appplot}


\begin{figure*}
	\includegraphics[width=3in]{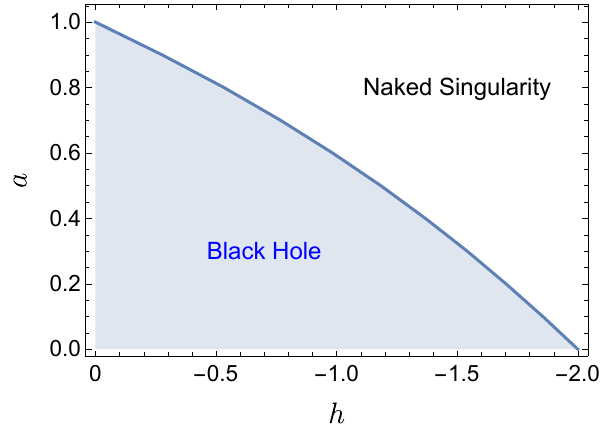}\,\,\,\,
	\includegraphics[width=3in]{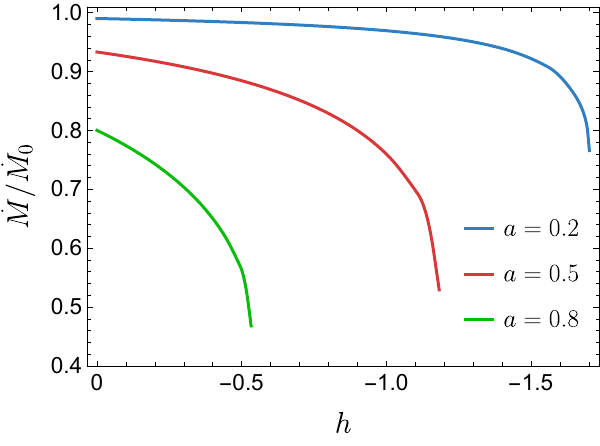}
	\caption{\textbf{Left:} Parameter space for rotating Horndeski black holes. The blue and white regions represent black hole solutions and naked singularity solutions, respectively, while the solid blue line corresponds to the extremal black hole solutions where the inner and outer horizons coincide. \textbf{Right:} Normalized accretion rate as a function of $h$, under different spin values. The disk thickness parameter has been set to $\sigma=0.2$.} 
	\label{parah}
\end{figure*}

\begin{figure*}[htbp]
	\centering
	\includegraphics[width=6.2in]{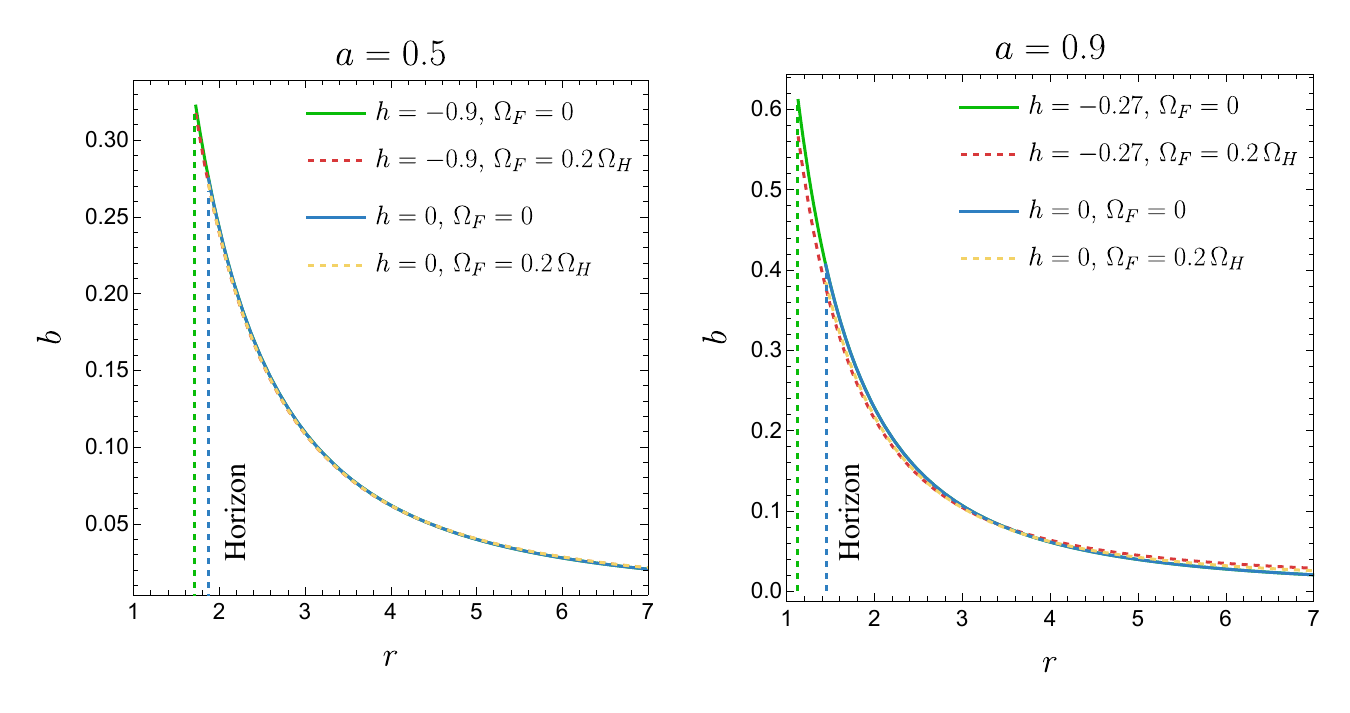}
	\caption{Variation of the magnetic-field strength $b=\sqrt{b^{\mu}b_{\mu}}$ with radius in the equatorial plane for different black hole parameters and field-line rotation rates.}
	\label{fig:bwithr}
\end{figure*}

\begin{figure*}[htbp]
	\centering
	\includegraphics[width=6.4in]{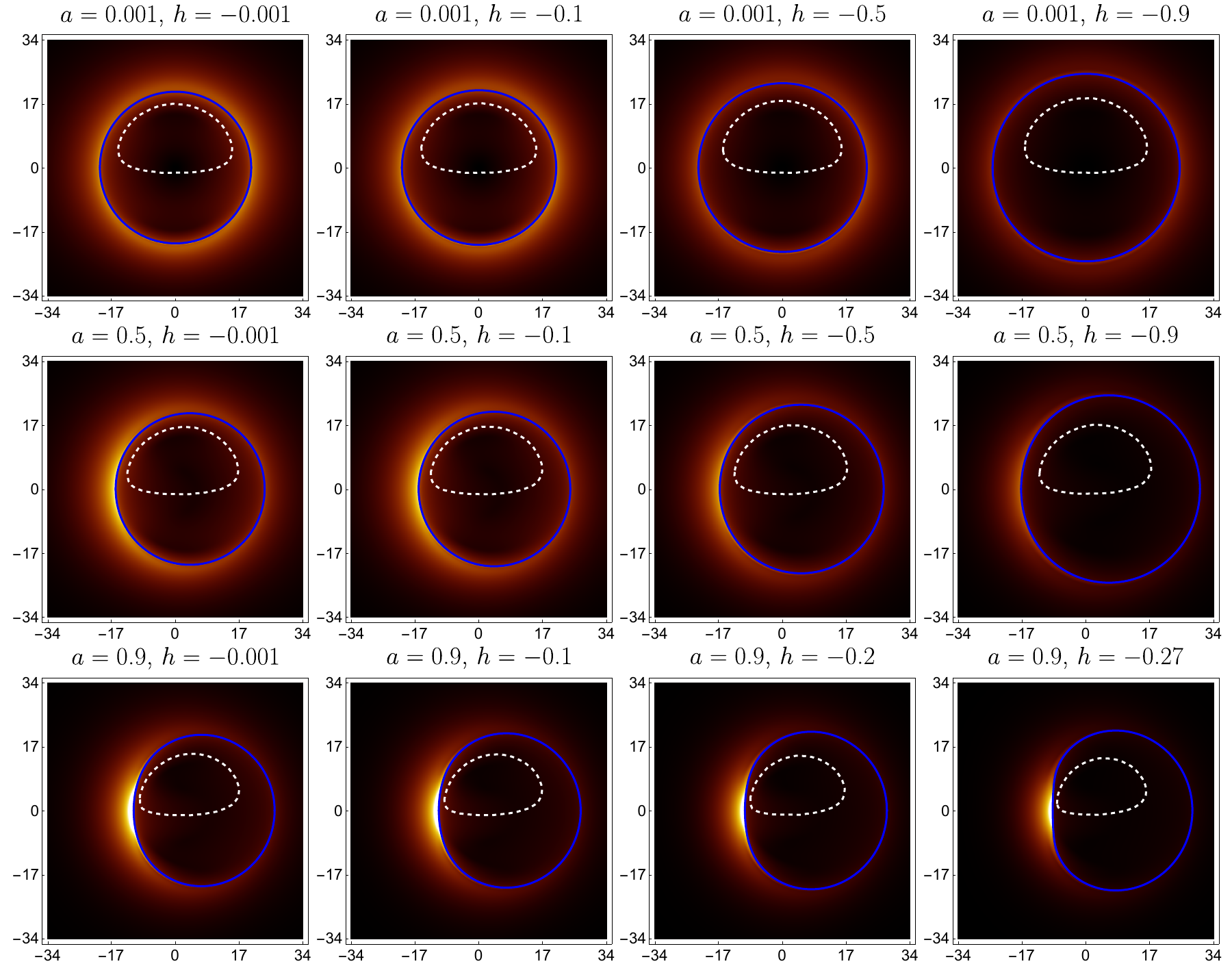}\\
     \quad \quad \,\,\, \includegraphics[width=6.2in]{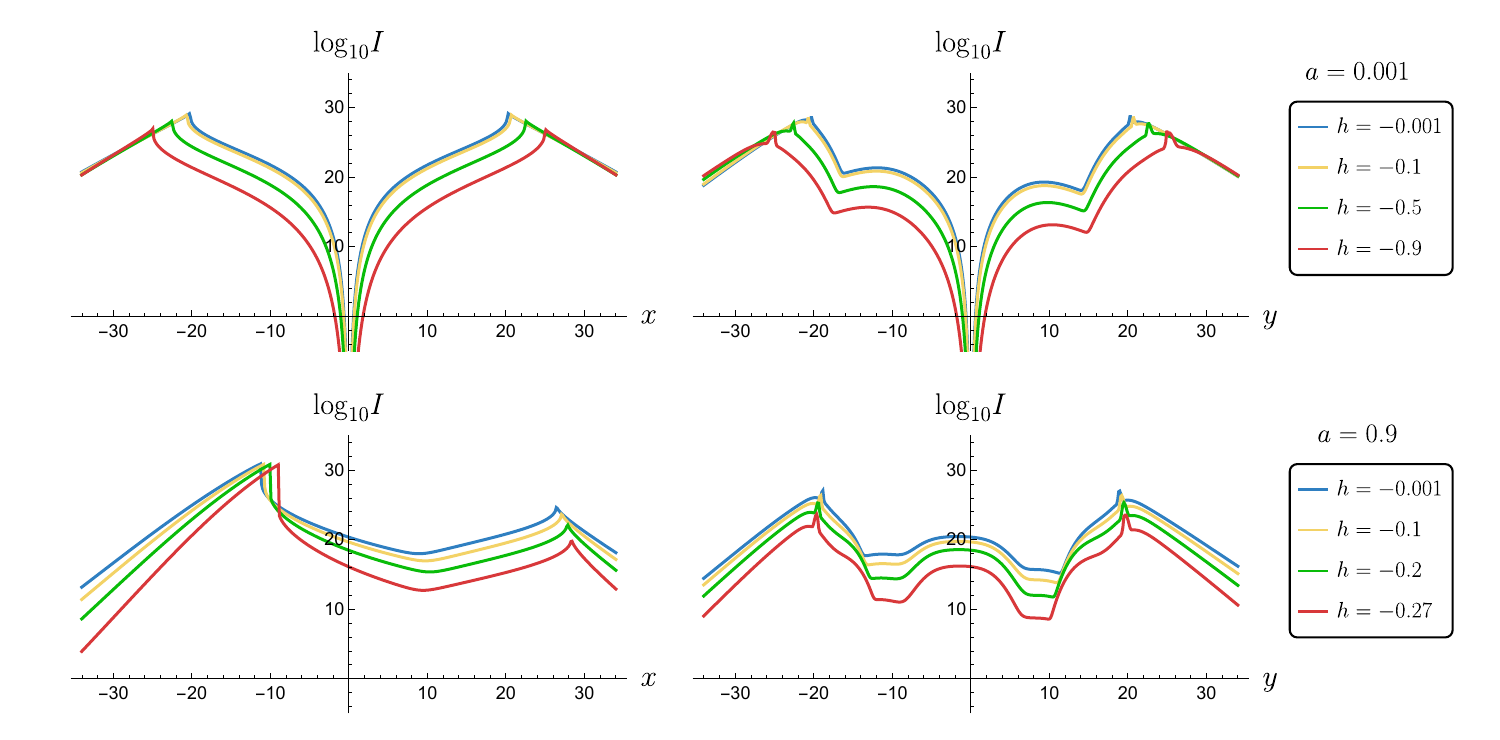}
	\caption{Imaging results of the thick accretion disk with a free-falling flow ($E=\mu$) around a rotating Horndeski black hole viewed at an inclination angle of $\theta_o=80^\circ$. 
The solid blue curve marks the critical curve, whereas the dashed white curve delineates the inner shadow boundary. 
The disk parameters are $z = 20, \sigma=0.2$ and $\Omega_F=0.2\,\Omega_H$.
The bottom two rows display the intensity profiles along the $x$-axis and $y$-axis for different black hole parameters.}
	\label{fig80}
\end{figure*}

\end{document}